\documentclass[aps,prc,twocolumn,twoside,superscriptaddress,nofootinbib,preprintnumbers,10pt,floatfix]{revtex4-2}
\usepackage{graphicx}
\usepackage{xcolor}
\usepackage{amsmath,amsfonts,amssymb}
\usepackage{physics}
\usepackage{hyperref}
\hypersetup{
    colorlinks=true,
    linkcolor=red,
    citecolor=blue,
    urlcolor=magenta
}
\usepackage{orcidlink}

\begin{document}

\title{Probing jet-medium interactions via jet substructure observables in relativistic heavy-ion collisions}

\author{Xiang-Pan Duan\orcidlink{0009-0001-0471-8832}}
\email{xpduan20@fudan.edu.cn}
\affiliation{Key Laboratory of Nuclear Physics and Ion-beam Application~(MOE), Institute of Modern Physics, Fudan University, Shanghai $200433$, China}
\affiliation{Shanghai Research Center for Theoretical Nuclear Physics, NSFC and Fudan University, Shanghai $200438$, China}
\affiliation{Instituto Galego de F\'isica de Altas Enerx\'ias IGFAE, Universidade de Santiago de Compostela, E-15782 Galicia, Spain}

\author{Tan Luo}
\email{luotan@hnu.edu.cn}
\affiliation{School of Physics and Electronics, Hunan University, Changsha 410082, China}

\author{Guo-Liang Ma\orcidlink{0000-0002-7002-8442}}
\email{glma@fudan.edu.cn}
\affiliation{Key Laboratory of Nuclear Physics and Ion-beam Application~(MOE), Institute of Modern Physics, Fudan University, Shanghai $200433$, China}
\affiliation{Shanghai Research Center for Theoretical Nuclear Physics, NSFC and Fudan University, Shanghai $200438$, China}


\begin{abstract}
We present a comprehensive study of jet substructure observables in $pp$ and PbPb collisions at $\sqrt{s_{NN}} = 5.02$~TeV using a multi-phase transport model. To suppress background contamination, the constituent subtraction method is employed for both PbPb and smeared $pp$ events. The jet splitting momentum fraction ($z_g$) and the groomed jet mass to the ungroomed jet transverse momentum ($M_g / p_{T,\text{jet}}$) are reconstructed using the Soft Drop algorithm with two grooming parameter settings. With $z_\text{cut} = 0.1$ and $\beta = 0.0$, a slight modification in the $z_g$ distribution is observed in central PbPb collisions, whereas a pronounced enhancement in the high $M_g / p_{T,\text{jet}}$ region is found, particularly at low $p_{T,\text{jet}}$ and in more central events. A detailed analysis of the dynamical evolution stages reveals that this enhancement primarily originates from jet-medium interactions, whereas the contributions from hadronization and hadronic rescatterings are largely mitigated by the grooming procedure. In contrast, under a stronger grooming condition ($z_\text{cut} = 0.5$, $\beta = 1.5$), no significant changes in $M_g / p_{T,\text{jet}}$ are observed, indicating that the medium-induced modifications are predominantly associated with large-angle scattering.
\end{abstract}

\maketitle

\section{Introduction}\label{sec:intro}

Relativistic heavy-ion collisions at the Relativistic Heavy Ion Collider (RHIC)~\cite{BRAHMS:2004adc,PHOBOS:2004zne,PHENIX:2004vcz,STAR:2005gfr,Eskola:2008ca,Luo:2017faz,Shen:2020mgh,Chen:2024aom} and the Large Hadron Collider (LHC)~\cite{ATLAS:2010isq,ALICE:2010yje,CMS:2012aa,Song:2017wtw,Shou:2024uga} create an extremely hot and dense nuclear matter referred to as the quark-gluon plasma (QGP)~\cite{Gyulassy:2004zy,Jia:2022ozr,He:2023zin}. In the early stage of the collisions, hard scatterings governed by quantum chromodynamics (QCD) produce high-energy partons with large virtuality. These partons subsequently undergo parton showering and evolve into collimated sprays of hadrons, referred to as jets~\cite{Sterman:1977wj,Feynman:1978dt,Duan:2025ngi}. As these energetic partons traverse the QGP medium, they undergo multiple interactions, including medium-induced gluon radiation and elastic scattering, resulting in energy loss and transverse momentum broadening. This phenomenon is known as jet quenching~\cite{Gyulassy:1990ye,Wang:1992qdg,Baier:1996sk,Salgado:2002cd,Salgado:2003gb,Armesto:2005iq,Casalderrey-Solana:2007knd,Qin:2015srf,Cao:2020wlm}, which leads to observable modifications in jet properties. Currently, jet quenching has become a widely used probe for accessing the transport properties of QGP. Measurements of jet observables, such as nuclear modification factor~\cite{STAR:2003fka,Eskola:2004cr,Cao:2017hhk,ATLAS:2018gwx,He:2018xjv,Zhao:2021vmu,Adhya:2021kws,Zhang:2024owr}, di-/$\gamma$-/Z-jet momentum imbalance~\cite{Qin:2010mn,Ma:2013pha,Chen:2016cof,Chen:2018fqu,Chen:2020kex,Li:2024uzk}, and jet transport coefficient~\cite{Liu:2015vna,Luo:2023nsi,Xie:2024xbn}, provide valuable insight into parton energy loss mechanisms and the characteristics of QGP medium.

In recent years, jet substructure observables have emerged as powerful tools to probe the jet-medium interactions. By examining the substructure of reconstructed jets, such as jet splitting fraction momentum ($z_g$)~\cite{CMS:2017qlm,STAR:2020ejj,ALICE:2021mqf,Wang:2022yrp}, jet splitting radius ($r_g$)~\cite{STAR:2020ejj,ALICE:2021mqf}, and groomed jet mass ($M_g$)~\cite{CMS:2018fof,STAR:2021lvw,ALICE:2024jtb}, these observables provide direct access to the mechanisms by which QGP modifies parton shower evolution. Jet grooming algorithms have been developed to remove the soft components, thereby isolating the hard components associated with the initial parton shower and finding the corresponding pair of subjets from the reconstructed jets.

A key motivation for studying groomed jet substructure lies in its sensitivity to jet-medium interactions in relativistic heavy-ion collisions. Both experimental measurements and theoretical predictions have reported diverse trends across different kinematic regimes and grooming conditions. For instance, as measured by CMS experiment~\cite{CMS:2017qlm}, the $z_g$ distribution in central PbPb collisions at high jet transverse momentum ($p_{T,\text{jet}} > 140$~GeV) exhibits a slight shift toward asymmetric splittings compared to peripheral PbPb and $pp$ collisions. This observation is consistent with predictions from several theoretical models, including JEWEL~\cite{Milhano:2017nzm} and SCET~\cite{Chien:2016led}. At lower jet transverse momentum ($60 < p_{T,\text{jet}} < 100$~GeV), ALICE experiment~\cite{ALICE:2021mqf} reports no significant modification of the $z_g$ distribution under strong grooming conditions, in line with results from JETSCAPE~\cite{JETSCAPE:2023hqn}. Interestingly, the higher twist formalism~\cite{Chang:2017gkt} predicts the strongest $z_g$ modification at intermediate jet energies, with weaker effects at both lower and higher energies.

The groomed jet mass has emerged as a sensitive observable for probing jet-medium interactions in relativistic heavy-ion collisions. Measurements by CMS experiment~\cite{CMS:2018fof} show a significant enhancement in high $M_g / p_{T,\text{jet}}$ region in central PbPb collisions, consistent with predictions from JEWEL~\cite{Milhano:2022kzx} and LBT~\cite{Luo:2021iay} models that incorporate medium response effects. In contrast, ALICE experiment reports a narrowing of the groomed jet mass distribution in the most central PbPb collisions compared to $pp$ collisions~\cite{ALICE:2024jtb}, in agreement with results from Hybrid model~\cite{Casalderrey-Solana:2014bpa} without elastic Moli\`{e}re scattering~\cite{DEramo:2018eoy}. Notably, when elastic Moli\`{e}re scattering is included in Hybrid model, a hint of enhancement at high $M_g$ region is observed, attributed to increased contributions from large-angle jet constituents. Similar behaviors are also observed in the ungroomed jet mass distributions~\cite{ALICE:2024jtb}, which are influenced by both perturbative and non-perturbative QCD~\cite{Duan:2023gmp}. Our previous study~\cite{Duan:2023gmp} demonstrated that the ungroomed jet mass is strongly affected by hadronization and hadronic rescatterings that can be substantially suppressed through the application of grooming techniques. A systematic investigation of jet substructure observables across different dynamic stages of jet evolution thus provides a valuable approach to understanding jet-medium interactions in relativistic heavy-ion collisions.

Measurements of $z_g$ and $M_g / p_{T,\text{jet}}$ serve as probes of jet quenching effects and provide valuable insight into the properties of QGP. However, interpreting these observables is complicated by the interplay between jet energy loss, medium response, and background fluctuations. In this context, theoretical modeling becomes crucial for disentangling these competing effects. In this work, we employ a multi-phase transport (AMPT) model to investigate the modification of jet substructure in $pp$ and PbPb collisions at $\sqrt{s_{NN}} = 5.02$~TeV. We focus on the $z_g$ and $M_g / p_{T,\text{jet}}$ observables and compare our results with experimental measurements from CMS~\cite{CMS:2017qlm,CMS:2018fof}. We further explore the dependence of these observables on jet transverse momentum, event centrality, and different dynamic stages of jet evolution, both with and without partonic interactions, to gain deeper insight into the jet substructure modifications.

The paper is organized as follows. In Sec.~\ref{sec:method}, we describe the AMPT framework, the jet reconstruction procedure, and the implementation of Soft Drop grooming. Section~\ref{sec:result-zg} presents the results on the $z_g$. In Sec.~\ref{sec:result-Mg}, we report the results of $M_g / p_{T,\text{jet}}$ and analyze the dependence on dynamical evolution stages. Finally, a summary is given in Sec.~\ref{sec:summary}.

\section{Methodology}\label{sec:method}

\subsection{The AMPT model}

The AMPT model with the string melting mechanism, widely used in relativistic heavy-ion collision studies, comprises four main stages~\cite{Lin:2004en,Lin:2021mdn}: initial conditions, parton cascade, hadronization, and hadronic rescatterings.

(1) \textit{Initial conditions.} The heavy ion jet interaction generator (HIJING) model~\cite{Wang:1991hta,Gyulassy:1994ew} provides the initial conditions for $pp$, $p$A, and AA collisions. Within the string melting mechanism, the primary interactions comprise two components: a soft contribution modeled by the Lund string fragmentation~\cite{Sjostrand:1993yb,Andersson:1983jt,Andersson:1983ia}, and a hard contribution from minijet production. The differential cross section for minijet production is computed using the perturbative QCD factorization:
\begin{equation}
    \frac{\dd \sigma^{cd}_{\rm \rm{jet}}}{\dd p_T^2 \dd y_1 \dd y_2} = K \sum_{a,b} x_1 f_{a}(x_1,Q^2) x_2 f_{b}(x_2,Q^2) \frac{\dd \hat{\sigma}^{ab \rightarrow cd}}{\dd \hat{t}},
\end{equation}
where $p_T$ is the transverse momentum of the produced minijet parton, $y_1$ and $y_2$ denote the rapidities of the final produced partons $c$ and $d$, the factor $K$ accounts for higher-order corrections beyond leading order (LO), and $x_1$, $x_2$ are the momentum fractions of the incoming partons $a$ and $b$. The parton distribution functions (PDFs) $f_{a}(x_1, Q^2)$ and $f_{b}(x_2, Q^2)$ follow the Duke-Owens parametrization~\cite{Duke:1983gd} with the factorization scale $Q^2$. The parton cross section $\dd \hat{\sigma}^{ab \rightarrow cd} / \dd \hat{t}$ is determined by the LO matrix elements squared $|\mathcal{M}_{ab \rightarrow cd}|^2$, expressed in terms of the Mandelstam variables $\hat{s}$, $\hat{t}$, and $\hat{u}$. A jet-triggering technique is employed in HIJING to generate dijet events, incorporating hard scattering processes such as $qq' \rightarrow qq'$, $qq \rightarrow qq$, $q\bar{q} \rightarrow q'\bar{q'}$, $q\bar{q} \rightarrow q\bar{q}$, $gq \rightarrow gq$, $q\bar{q} \rightarrow gg$, $gg \rightarrow q\bar{q}$, and $gg \rightarrow gg$~\cite{Owens:1986mp,Sjostrand:1993yb}.

(2) \textit{Parton cascade.} The Zhang’s parton cascade (ZPC) model describes partonic interactions via two-body elastic scatterings~\cite{Zhang:1997ej}. The parton cross section for gluon-gluon scattering is evaluated at LO pQCD as
\begin{equation}
    \sigma_{gg} \approx \frac{9\pi \alpha_s^2}{2\mu^2},
\end{equation}
where $\alpha_s$ is the strong coupling constant and $\mu$ is the Debye screening mass. Varying $\mu$ allows adjustment of the parton cross section in simulations.

(3) \textit{Hadronization.} Hadronization is modeled using the quark coalescence mechanism~\cite{Lin:2004en}, which recombines the nearest two or three partons into mesons and baryons without considering their relative momentum. Three momentum conservation is satisfied in the coalescence process, and the hadron species are determined by the flavor and invariant mass of the coalescing partons.

(4) \textit{Hadronic rescatterings.} The relativistic transport (ART) model~\cite{Li:1995pra} simulates resonance decays and hadronic reactions, including both elastic and inelastic scatterings for baryon-baryon, baryon-meson, and meson-meson interactions.

In this work, AMPT model with the string melting mechanism is used to simulate $pp$ and PbPb collisions at $\sqrt{s_{NN}} = 5.02$~TeV. The parton cross section is set to 3~mb, in line with previous studies that have successfully described the collective flow observed in relativistic heavy-ion collisions~\cite{Ma:2014pva,Bzdak:2014dia,Bozek:2015swa,He:2015hfa,Tang:2023wcd,Duan:2024mwa}. The AMPT model incorporating jet-medium interactions has been widely utilized in the investigation of various jet observables, including dijet asymmetry~\cite{Ma:2013pha}, $\gamma$-jet imbalance~\cite{Ma:2013bia}, jet fragmentation functions~\cite{Ma:2013gga,Ma:2013yoa,Duan:2022bew,Feng:2024tmc}, jet shape~\cite{Ma:2013uqa}, jet anisotropies~\cite{Nie:2014pla}, jet transport coefficients~\cite{Zhou:2019gqk}, the redistribution of lost energy~\cite{Gao:2016ldo,Luo:2021hoo,Luo:2021voy,Luo:2024xog}, and jet mass~\cite{Duan:2023gmp}. For comparison, a parton cross section of 0~mb is also employed to provide a baseline scenario without jet-medium interactions.

\subsection{Jet reconstruction}

To measure jet substructure observables, jets are first reconstructed using the anti-$k_t$ algorithm~\cite{Cacciari:2008gp} with a radius parameter $R = 0.4$, as implemented in the FastJet package~\cite{Cacciari:2011ma}. In this analysis, jets are required to have pseudorapidity $|\eta_\text{jet}| < 1.3$. To facilitate direct comparison with CMS measurements~\cite{CMS:2017qlm,CMS:2018fof}, the masses of final-state particles are adjusted such that neutral hadrons are assigned zero mass, while charged hadrons are set to the charged pion mass, in accordance with the experimental setup used by CMS. This treatment has a non-negligible effect on the groomed jet mass distribution and is essential for a consistent comparison between simulation and data.

In PbPb collisions, the constituent subtraction method~\cite{Berta:2014eza} is employed to estimate the background, characterized by the background density $\rho$ and background mass density $\rho_m$~\cite{Cacciari:2011ma,Soyez:2012hv}. The estimation is based on clustering particles using the $k_t$ algorithm~\cite{Cacciari:2005hq} with $R = 0.4$. Massless ghost particles with an area of 0.005 are added in the $y$-$\phi$ plane to the event. In order to reduce contamination from hard jet fragments, the two $k_t$ clusters with the highest transverse momentum are excluded. The constituent subtraction is performed on a particle-by-particle level, enabling correction of both jet four-momentum and its substructure. This method is applied in PbPb events generated with the AMPT model using a jet-triggering technique. To avoid direct comparison between $pp$ and PbPb collisions, a smeared procedure similar to that used in the CMS experiment is adopted. Specifically, the background from PbPb collisions (without jet trigger) is embedded into $pp$ collisions with jet trigger to produce the so-called “smeared” $pp$ events. The constituent subtraction procedure in PbPb collisions is applied to the smeared $pp$ events to maintain consistency.

\begin{figure*}[htbp]
    \centering
    \includegraphics[width=0.42\textwidth]{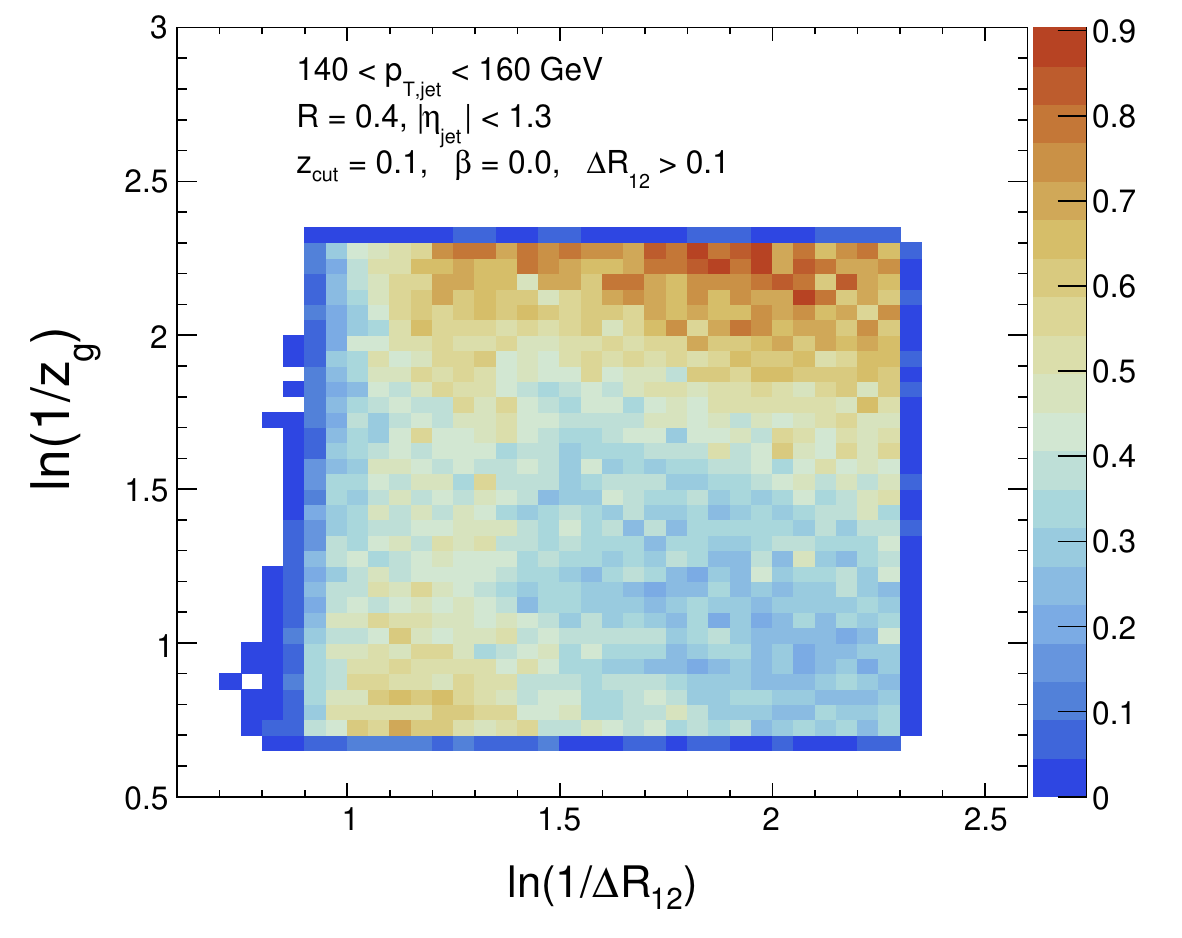}
    \includegraphics[width=0.42\textwidth]{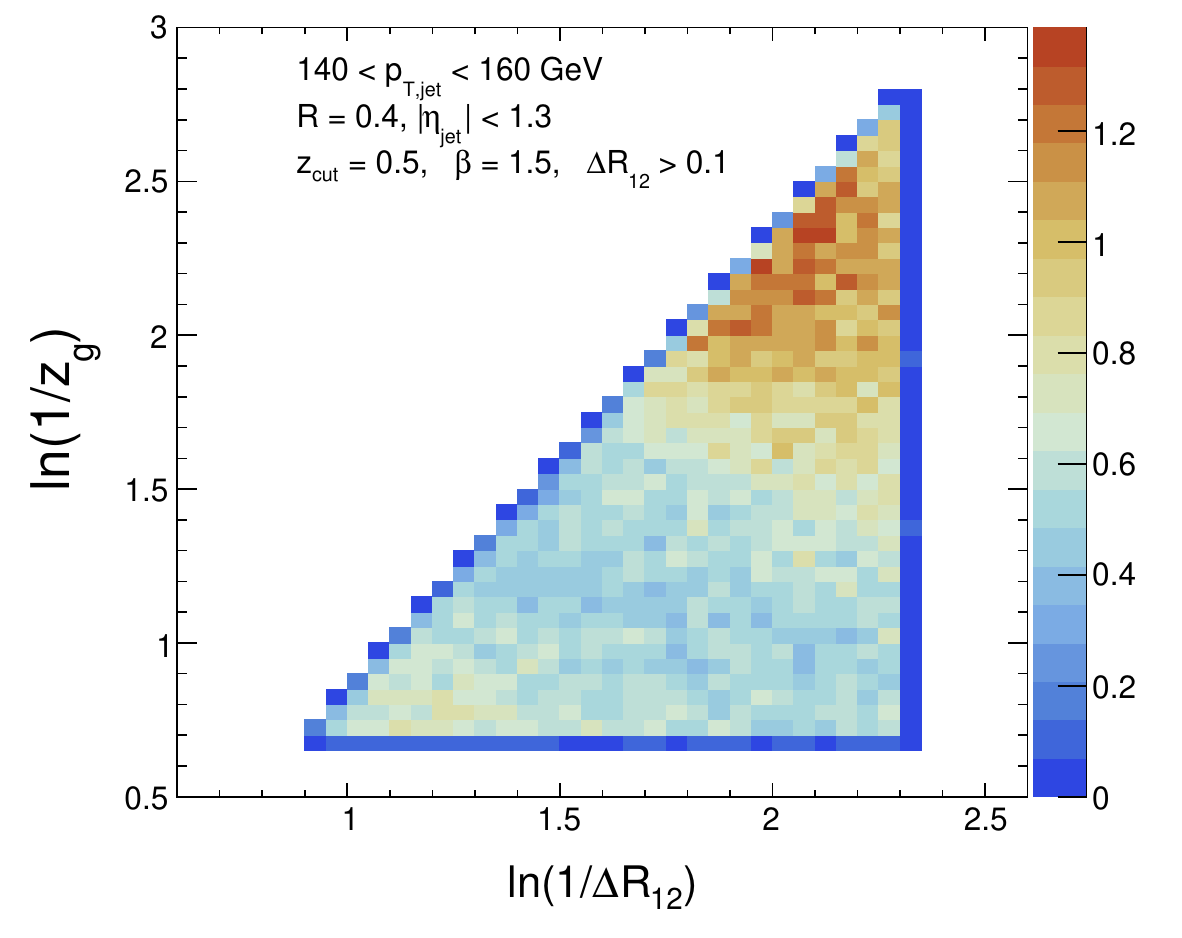}
    \caption{Comparison of two Soft Drop grooming parameter settings: $z_\text{cut} = 0.1$ and $\beta = 0.0$ (left panel) and $z_\text{cut} = 0.5$ and $\beta = 1.5$ (right panel), with an angular cut of $\Delta R_{12} > 0.1$, for jets in the transverse momentum range $140 < p_{T,\text{jet}} < 160$~GeV in $pp$ collisions at $\sqrt{s} = 5.02$~TeV.}
    \label{fig:pp-RgZg}
\end{figure*}

\subsection{Soft Drop}

In addition, jet grooming is performed using the Soft Drop algorithm~\cite{Larkoski:2014wba}, which removes the soft components and isolates the hard components of the jet, thereby enhancing sensitivity to the medium-induced modifications. The procedure reclusters jet constituents using the Cambridge-Aachen (C/A) algorithm~\cite{Dokshitzer:1997in} to form a pairwise clustering tree, followed by recursive declustering. At each declustering step, the jet is separated into two subjets by undoing the last C/A clustering stage. The subjets are required to satisfy the Soft Drop condition:
\begin{equation}\label{eq:sd}
    \frac{\min(p_{T,1}, p_{T,2})}{p_{T,1} + p_{T,2}} > z_\text{cut} (\frac{\Delta R_{12}}{R})^\beta,
\end{equation}
where $p_{T,1}$ and $p_{T,2}$ are the transverse momenta of the subjets, $\Delta R_{12} = \sqrt{(y_1-y_2)^2+(\phi_1-\phi_2)^2}$ is the distance between the subjets in the $y$-$\phi$ plane, and $R = 0.4$ is the jet radius. The parameters $z_\text{cut}$ and $\beta$ control the grooming procedure. If the Soft Drop condition is satisfied, the corresponding pair of subjets is retained as the final groomed jet. Otherwise, the subjet with higher $p_T$ is further declustered, and the procedure is repeated until the condition is satisfied. The final groomed subjets are used to measure the jet substructure properties.

In this study, groomed jet observables, including the jet splitting momentum fraction $z_g$~\cite{CMS:2017qlm} and the groomed jet mass to the ungroomed jet transverse momentum $M_g / p_{T,\text{jet}}$~\cite{CMS:2018fof}, are calculated using the AMPT model with a jet-triggering technique. The jet splitting momentum fraction $z_g$ is defined as
\begin{equation}\label{eq:zg}
    z_g = \frac{\min(p_{T,1}, p_{T,2})}{p_{T,1} + p_{T,2}},
\end{equation}
where $p_{T,1}$ and $p_{T,2}$ are the transverse momenta of the two subjets. This definition is identical to the left side of Eq.~\eqref{eq:sd} when the subjets satisfy the Soft Drop condition. The groomed jet mass $M_g$ is defined as
\begin{equation}\label{eq:Mg}
    M_g = \sqrt{(E_1 + E_2)^2 - (\Vec{p}_1 + \Vec{p}_2)^2},
\end{equation}
where $E_i$ and $\Vec{p}_i$ are the energy and three-momentum vectors of the subjets, respectively. To reconstruct jet substructure, two sets of Soft Drop grooming parameters are employed, each probing distinct regions of the subjet phase space in the Lund plane~\cite{Dreyer:2018nbf,Andrews:2018jcm}. The first setup, with ($z_\text{cut} = 0.1$ and $\beta = 0.0$), imposes a grooming condition based solely on the energy fraction between subjets to capture both jet core and peripheral modifications and is utilized for analyzing both $z_g$ and $M_g / p_{T,\text{jet}}$ observables. The second setup, which employs a stronger grooming condition ($z_\text{cut} = 0.5$, $\beta = 1.5$), reduces the contribution from large-angle subjets, thereby enhancing sensitivity to the jet core region. This configuration is employed exclusively in the $M_g / p_{T,\text{jet}}$ analysis. An additional angular requirement of $\Delta R_{12} > 0.1$ is imposed across all measurements to suppress unphysical contributions, following the CMS experimental procedure~\cite{CMS:2017qlm,CMS:2018fof}. Furthermore, the $z_g$ and $M_g / p_{T,\text{jet}}$ distributions are normalized by the number of groomed jets ($N_{g,\text{jet}}$). To facilitate a more direct comparison between the two grooming configurations, Fig.~\ref{fig:pp-RgZg} presents the joint distributions of $z_g$ and $\Delta R_{12}$ for jets that satisfy the Soft Drop condition defined in Eq.~\eqref{eq:sd}. The distributions include an angular cut of $\Delta R_{12} > 0.1$ and are normalized by $\frac{1}{N_{g,\text{jet}}} \frac{\dd^2 N_{g,\text{jet}}}{\dd\ln(1/\Delta R_{12})~\dd\ln(1/z_g)}$. The comparison is performed within the jet transverse momentum range of $140 < p_{T,\text{jet}} < 160$~GeV in $pp$ collisions at $\sqrt{s} = 5.02$~TeV using the AMPT model. Relative to the first setup ($z_\text{cut} = 0.1$, $\beta = 0.0$), the stronger grooming configuration ($z_\text{cut} = 0.5$, $\beta = 1.5$) exhibits a clearly depleted region in the upper-left part of the distribution, corresponding to a reduction in subjets with large angular separation. This indicates an increased focus on the jet core region.

\begin{figure*}[htbp]
    \centering
    \includegraphics[width=0.42\textwidth]{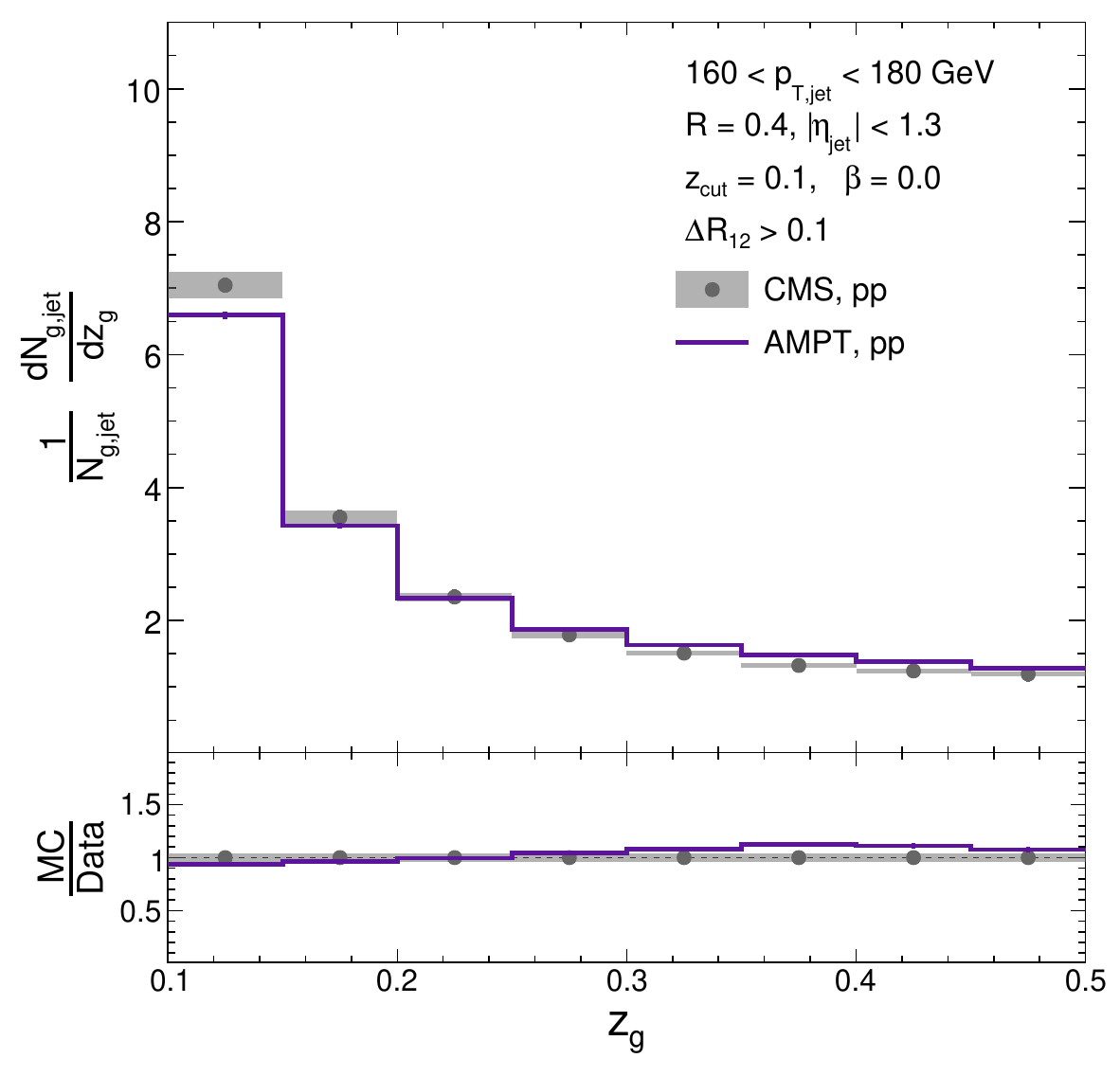}
    \includegraphics[width=0.42\textwidth]{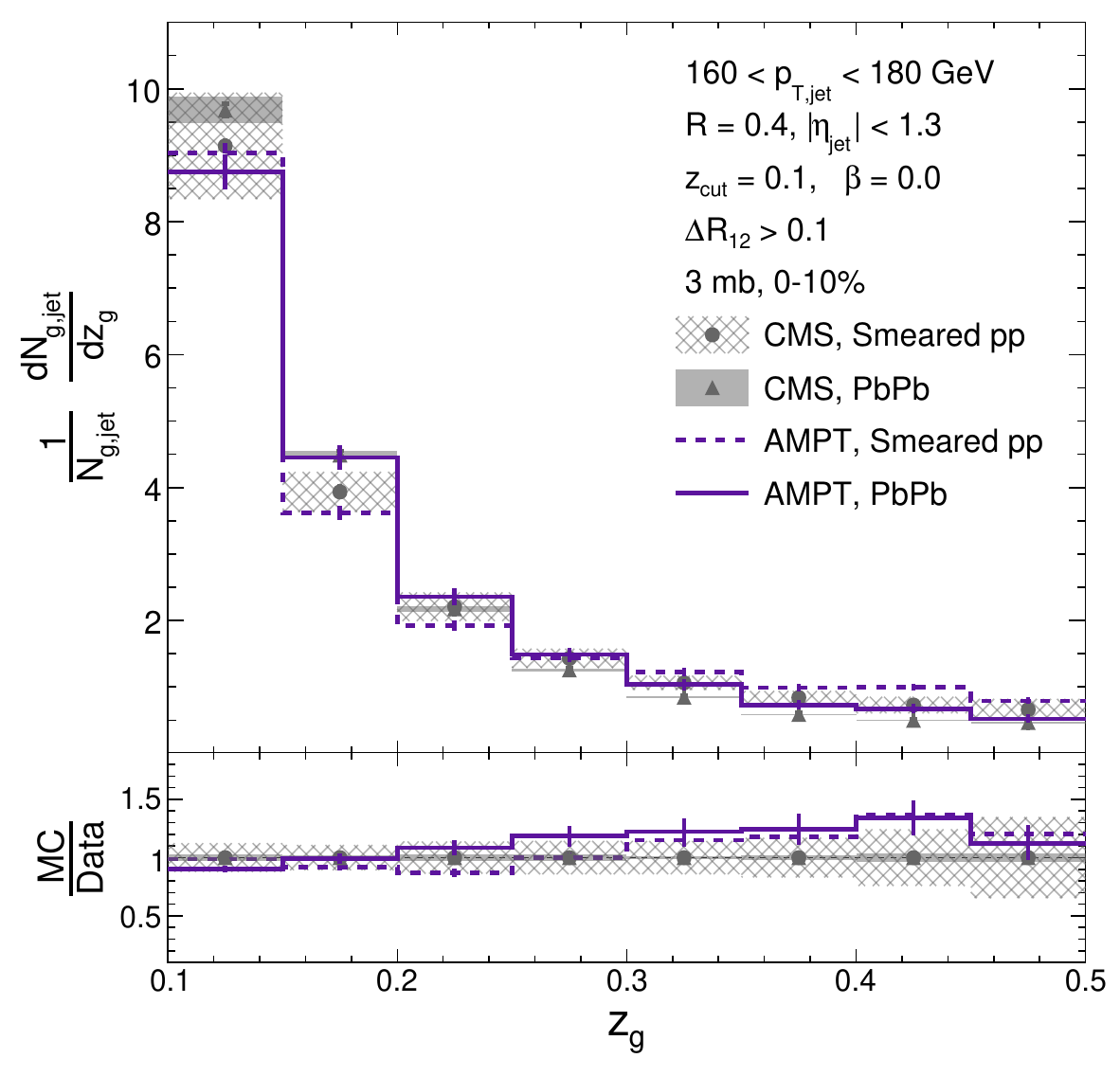}
    \caption{Distributions of the $z_g$ in $pp$ collisions (left panel) and in 0-10\% smeared $pp$ and PbPb collisions (right panel), within the jet transverse momentum range of $160 < p_{T,\text{jet}} < 180$~GeV at $\sqrt{s_{NN}} = 5.02$~TeV. The Soft Drop grooming parameters are set to $z_\text{cut} = 0.1$ and $\beta = 0.0$, with an angular cut of $\Delta R_{12} > 0.1$. Lines represent the AMPT model results, while data points correspond to CMS measurements~\cite{CMS:2017qlm}, with statistical uncertainties shown as error bars and systematic uncertainties as shaded bands.}
    \label{fig:pp-PbPb-Zg}
\end{figure*}

\begin{figure*}[htbp]
    \centering
    \includegraphics[width=0.42\textwidth]{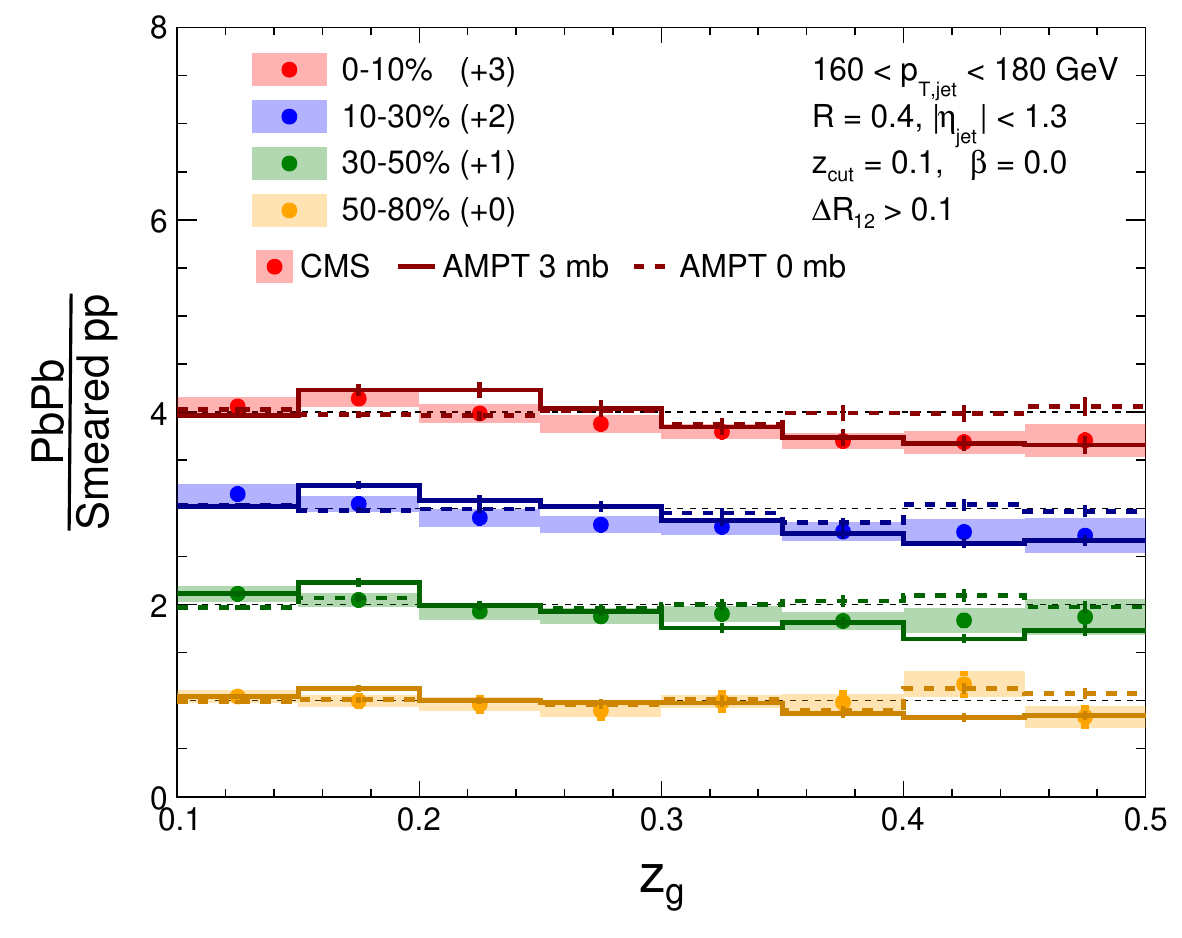}
    \includegraphics[width=0.42\textwidth]{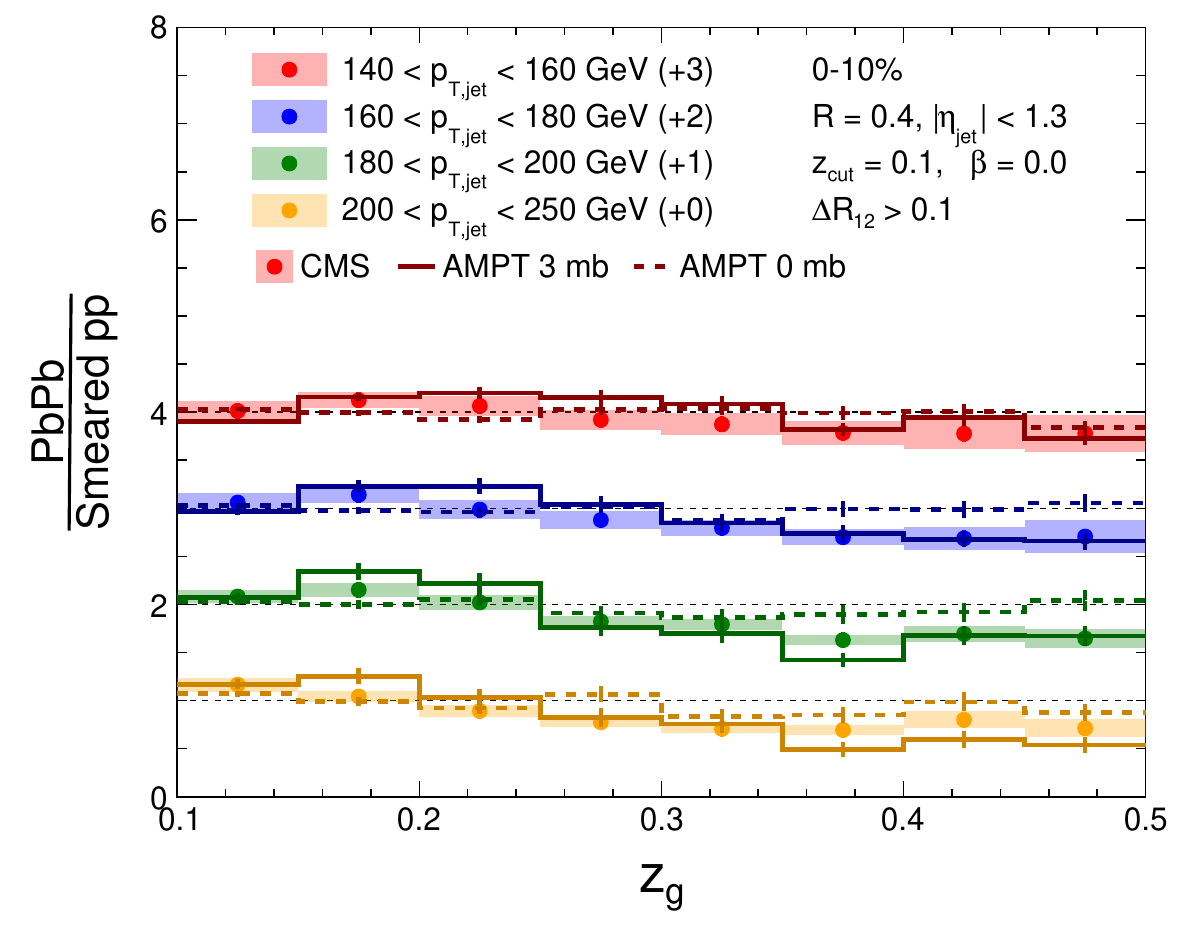}
    \caption{Ratios of the $z_g$ distributions between PbPb and smeared $pp$ collisions at $\sqrt{s_{NN}} = 5.02$~TeV, shown as a function of centrality (left panel) and jet transverse momentum $p_{T,\text{jet}}$ (right panel). The Soft Drop grooming parameters are set to $z_\text{cut} = 0.1$ and $\beta = 0.0$, with an angular cut of $\Delta R_{12} > 0.1$. Solid (3~mb) and dashed (0~mb) lines represent the AMPT model results, while data points correspond to CMS measurements~\cite{CMS:2017qlm}, with statistical uncertainties shown as error bars and systematic uncertainties as shaded bands.}
    \label{fig:PbPb-Zg-cen-pT}
\end{figure*}

\begin{figure*}[htbp]
    \centering
    \includegraphics[width=0.42\textwidth]{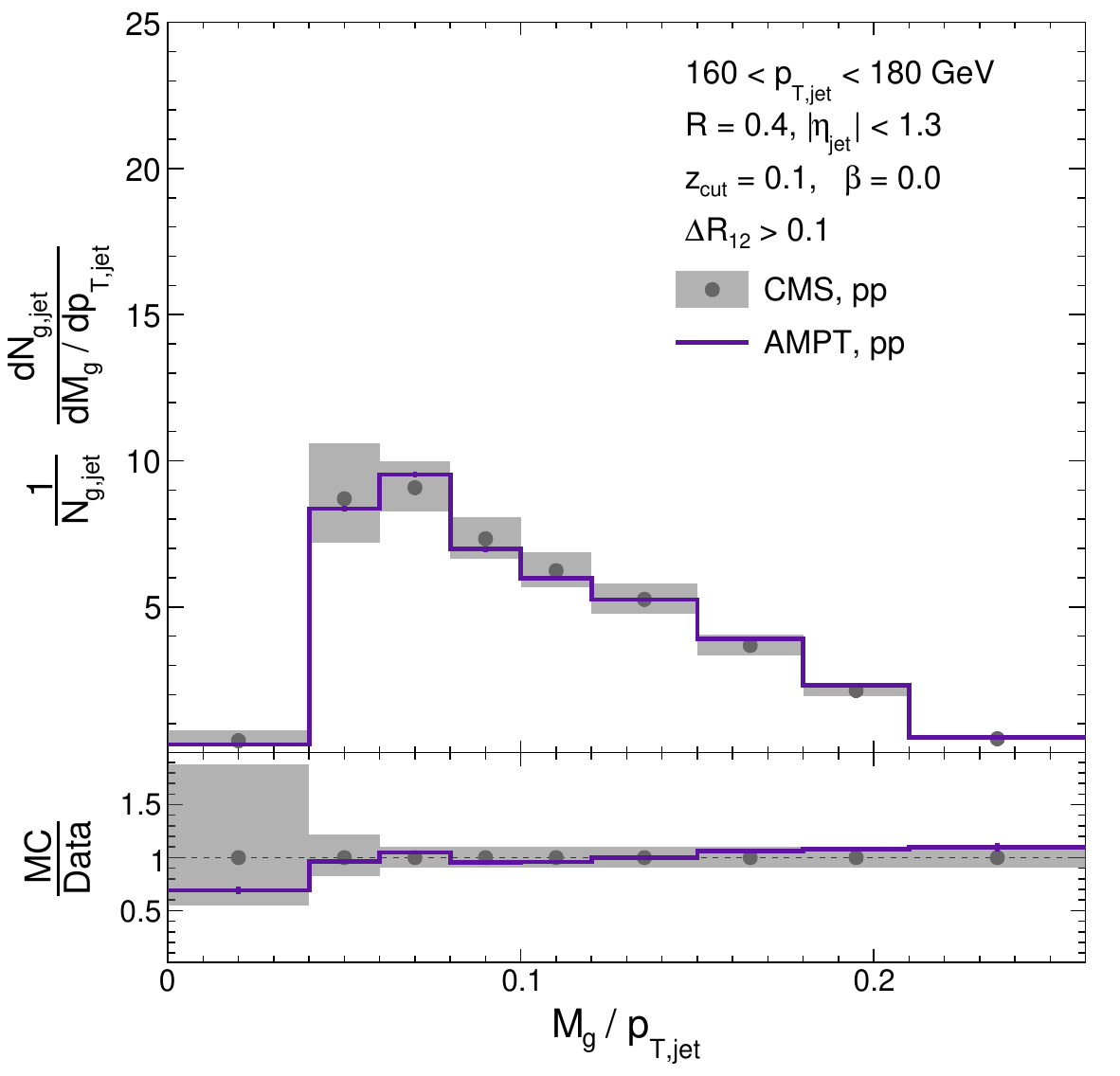}
    \includegraphics[width=0.42\textwidth]{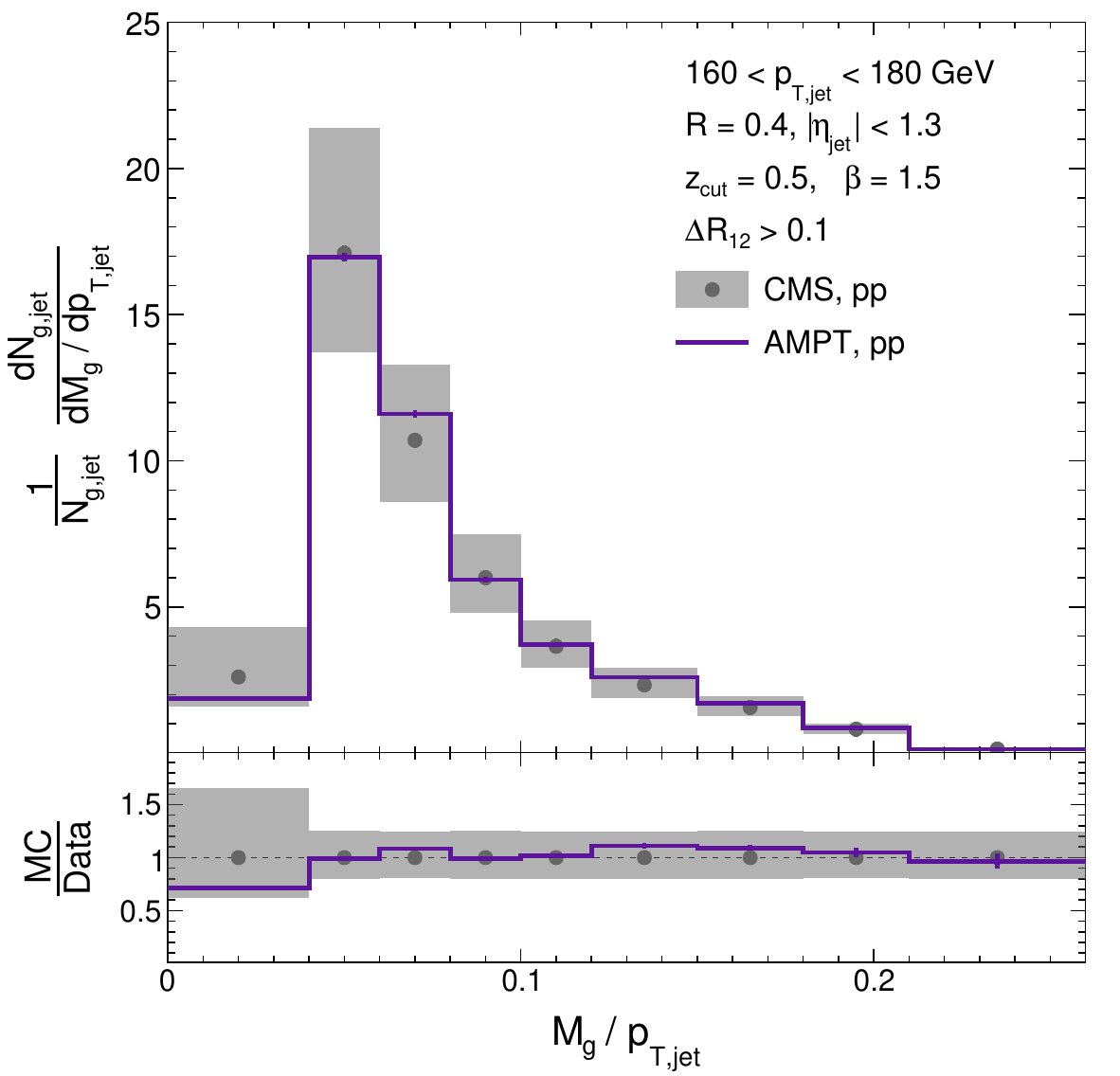}
    \includegraphics[width=0.42\textwidth]{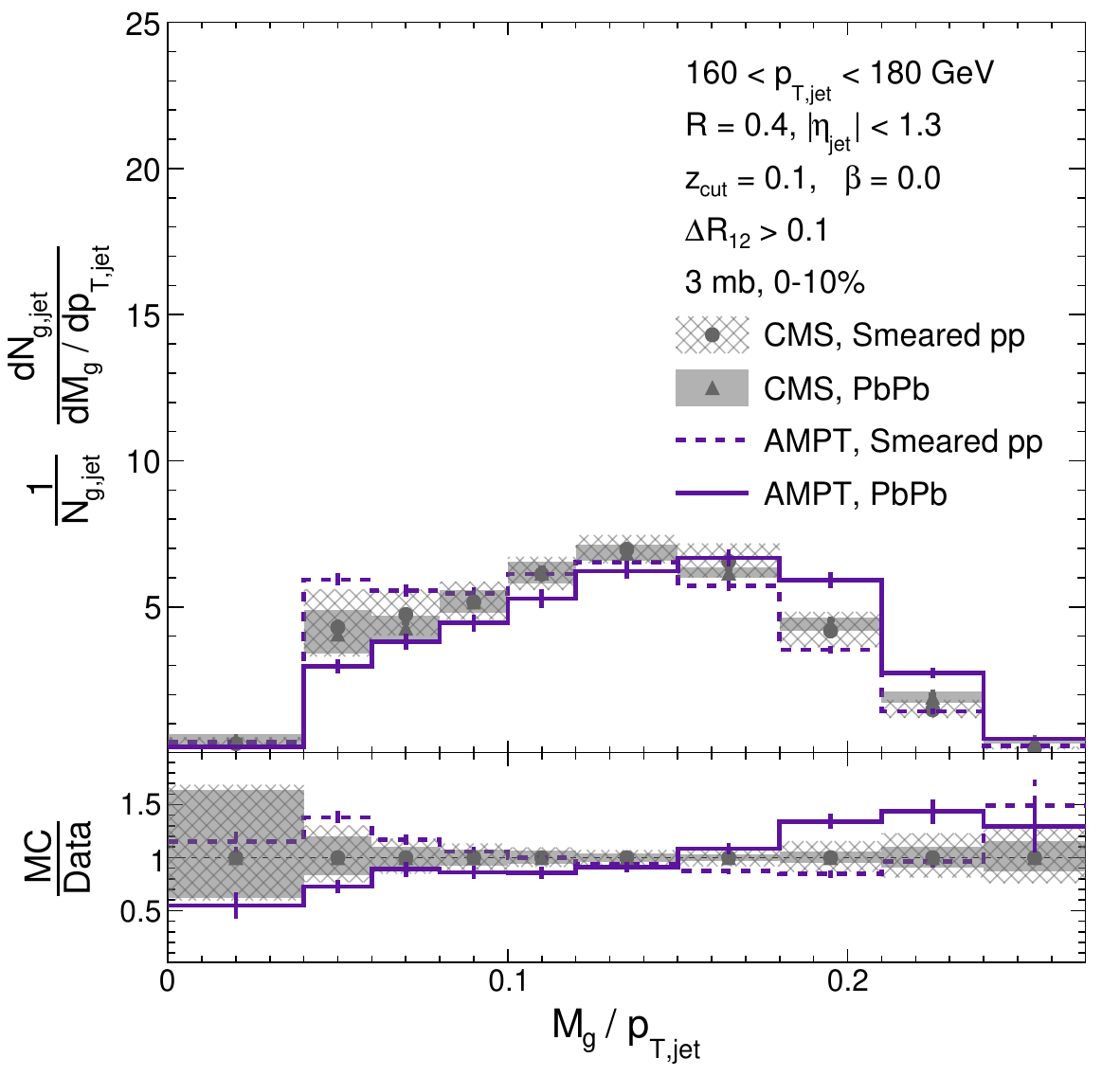}
    \includegraphics[width=0.42\textwidth]{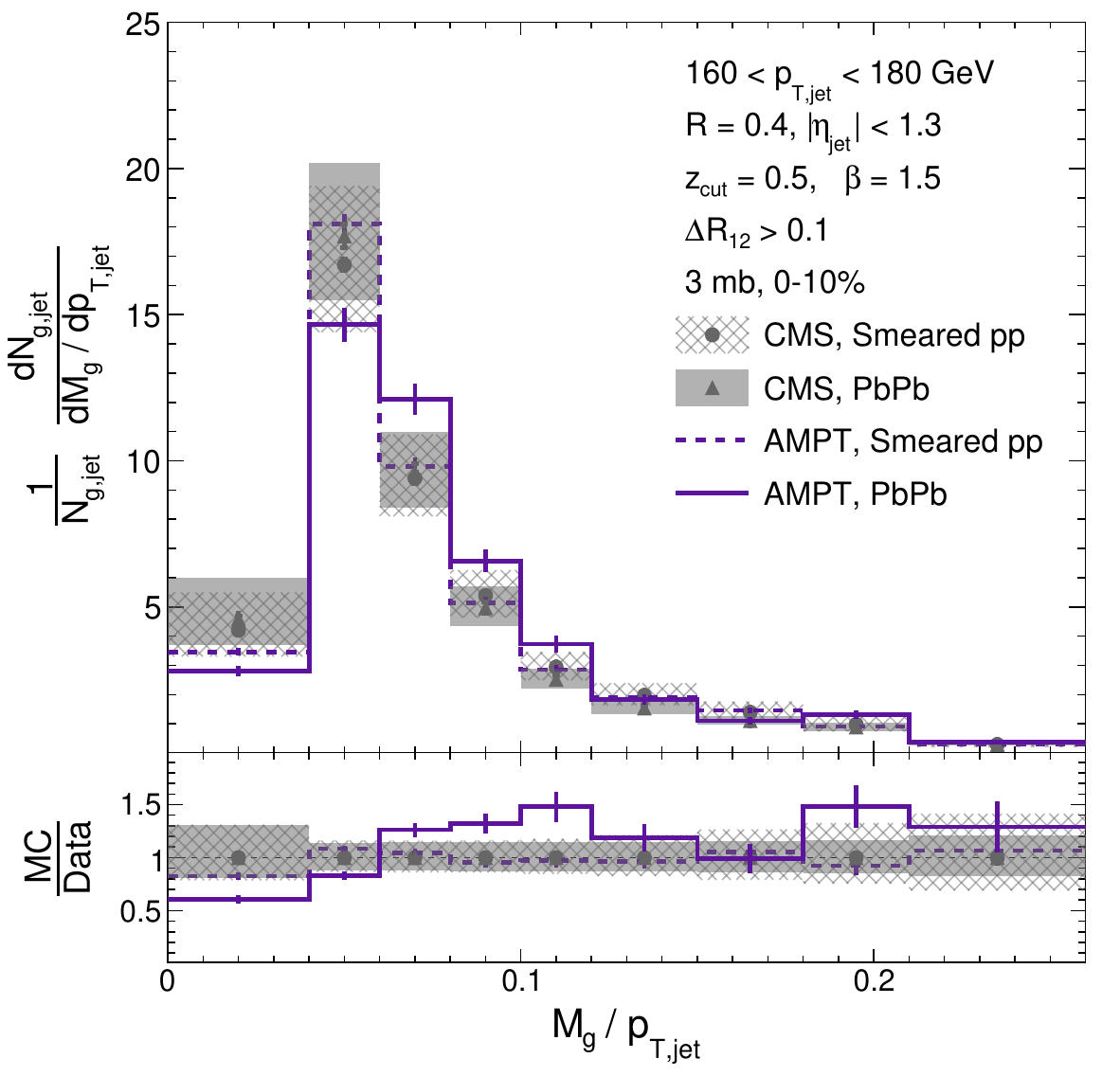}
    \caption{Distributions of the $M_g / p_{T,\text{jet}}$ in $pp$ collisions (top panels) and in 0-10\% smeared $pp$ and PbPb collisions (bottom panels) within the jet transverse momentum range of $160 < p_{T,\text{jet}} < 180$~GeV at $\sqrt{s_{NN}} = 5.02$~TeV. The Soft Drop grooming parameters are set to $z_\text{cut} = 0.1$, $\beta = 0.0$ (left panels) and $z_\text{cut} = 0.5$, $\beta = 1.5$ (right panels), with an angular cut of $\Delta R_{12} > 0.1$. Lines represent the AMPT model results, while data points correspond to CMS measurements~\cite{CMS:2018fof}, with statistical uncertainties shown as error bars and systematic uncertainties as shaded bands.}
    \label{fig:pp-PbPb-Mg}
\end{figure*}

\begin{figure*}[htbp]
    \centering
    \includegraphics[width=0.42\textwidth]{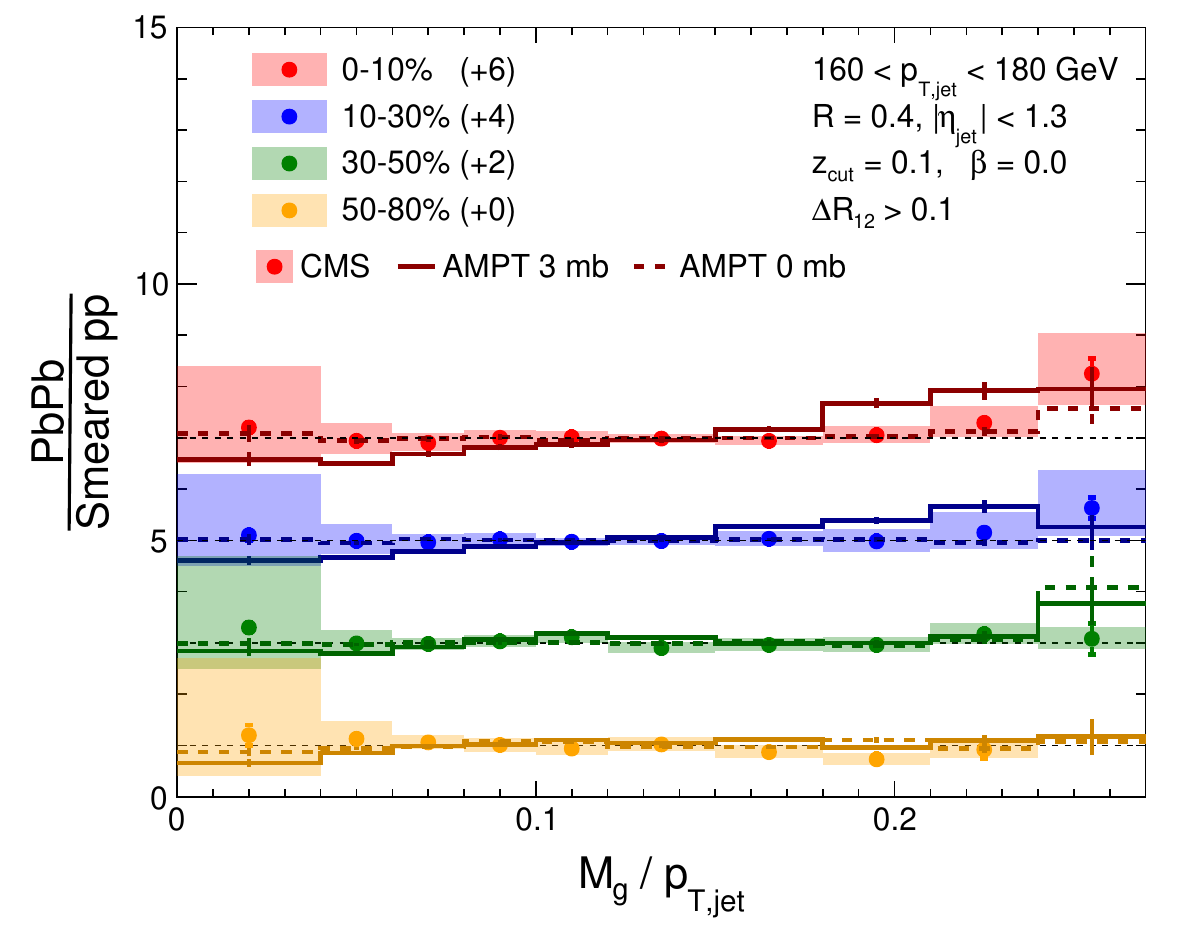}
    \includegraphics[width=0.42\textwidth]{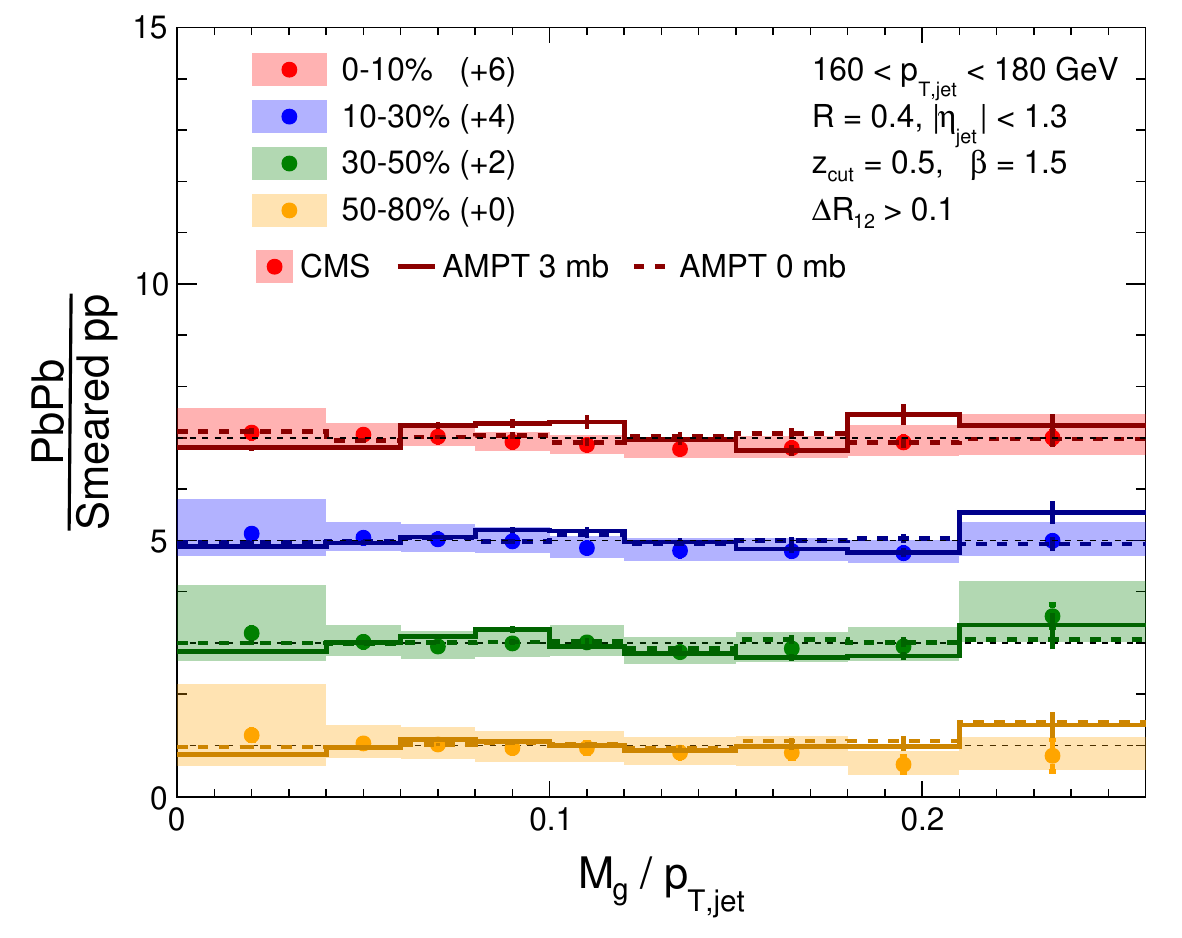}
    \caption{Ratios of the $M_g / p_{T,\text{jet}}$ distributions between PbPb and smeared $pp$ collisions in the jet transverse momentum range of $160 < p_{T,\text{jet}} < 180$~GeV at $\sqrt{s_{NN}} = 5.02$~TeV, shown as a function of centrality. The Soft Drop grooming parameters are set to $z_\text{cut} = 0.1$, $\beta = 0.0$ (left panel) and $z_\text{cut} = 0.5$, $\beta = 1.5$ (right panel), with an angular cut of $\Delta R_{12} > 0.1$. Solid (3~mb) and dashed (0~mb) lines represent the AMPT model results, while data points correspond to CMS measurements~\cite{CMS:2018fof}, with statistical uncertainties shown as error bars and systematic uncertainties as shaded bands.}
    \label{fig:PbPb-Mg-cen}
\end{figure*}

\begin{figure*}[htbp]
    \centering
    \includegraphics[width=0.42\textwidth]{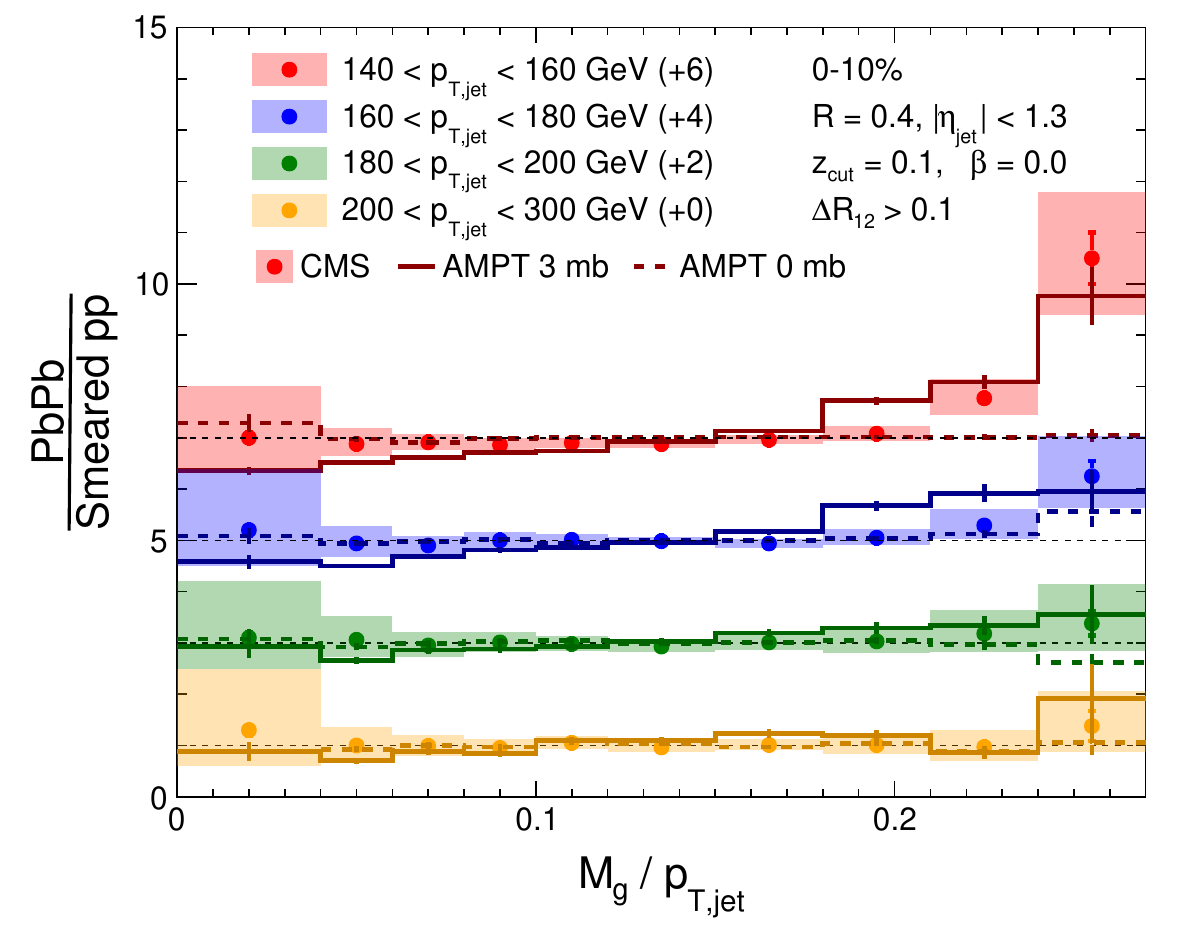}
    \includegraphics[width=0.42\textwidth]{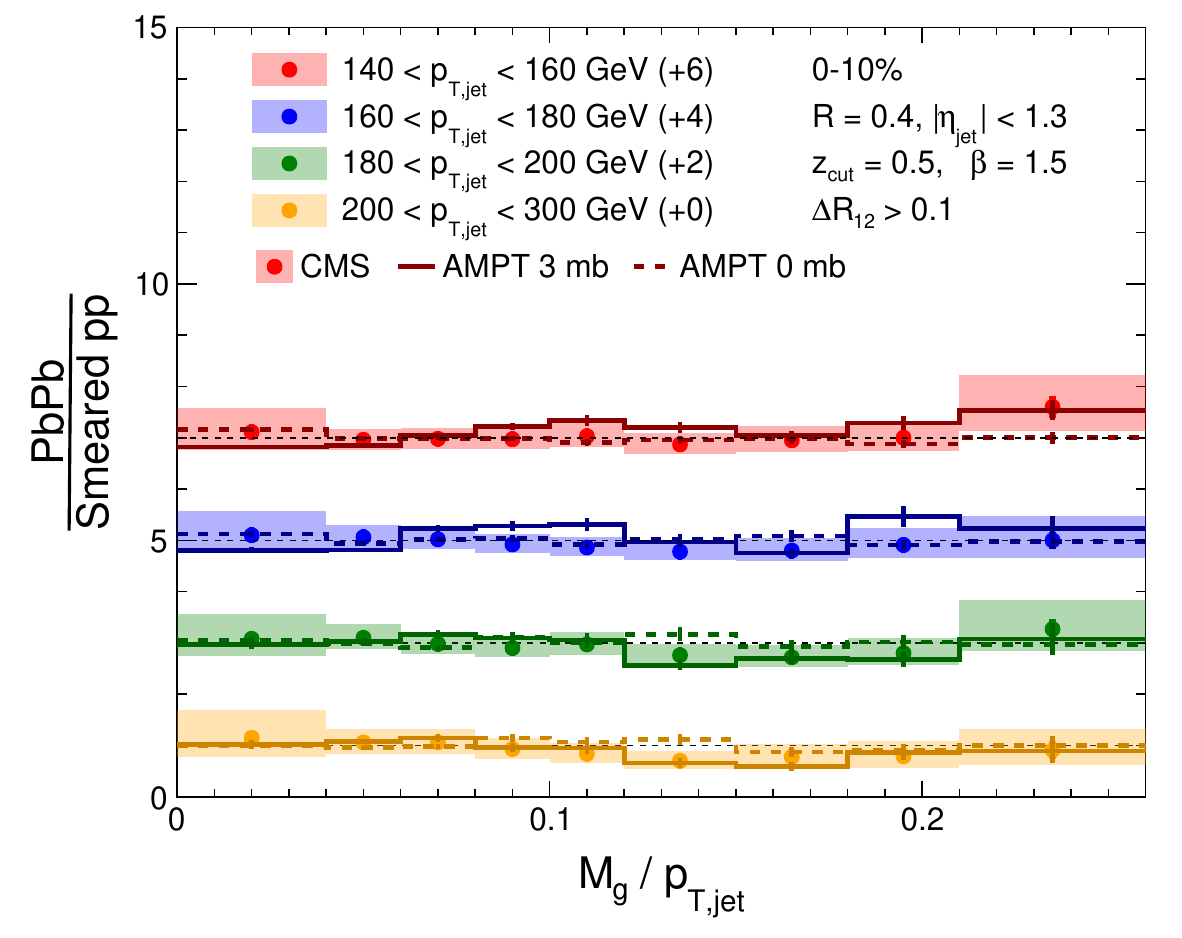}
    \caption{Ratios of the $M_g / p_{T,\text{jet}}$ distributions between 0-10\% PbPb and smeared $pp$ collisions at $\sqrt{s_{NN}} = 5.02$~TeV, shown as a function of jet transverse momentum $p_{T,\text{jet}}$. The Soft Drop grooming parameters are set to $z_\text{cut} = 0.1$, $\beta = 0.0$ (left panel) and $z_\text{cut} = 0.5$, $\beta = 1.5$ (right panel), with an angular cut of $\Delta R_{12} > 0.1$. Solid (3~mb) and dashed (0~mb) lines represent the AMPT model results, while data points correspond to CMS measurements~\cite{CMS:2018fof}, with statistical uncertainties shown as error bars and systematic uncertainties as shaded bands.}
    \label{fig:PbPb-Mg-pT}
\end{figure*}

\begin{figure*}[htbp]
    \centering
    \includegraphics[width=0.42\textwidth]{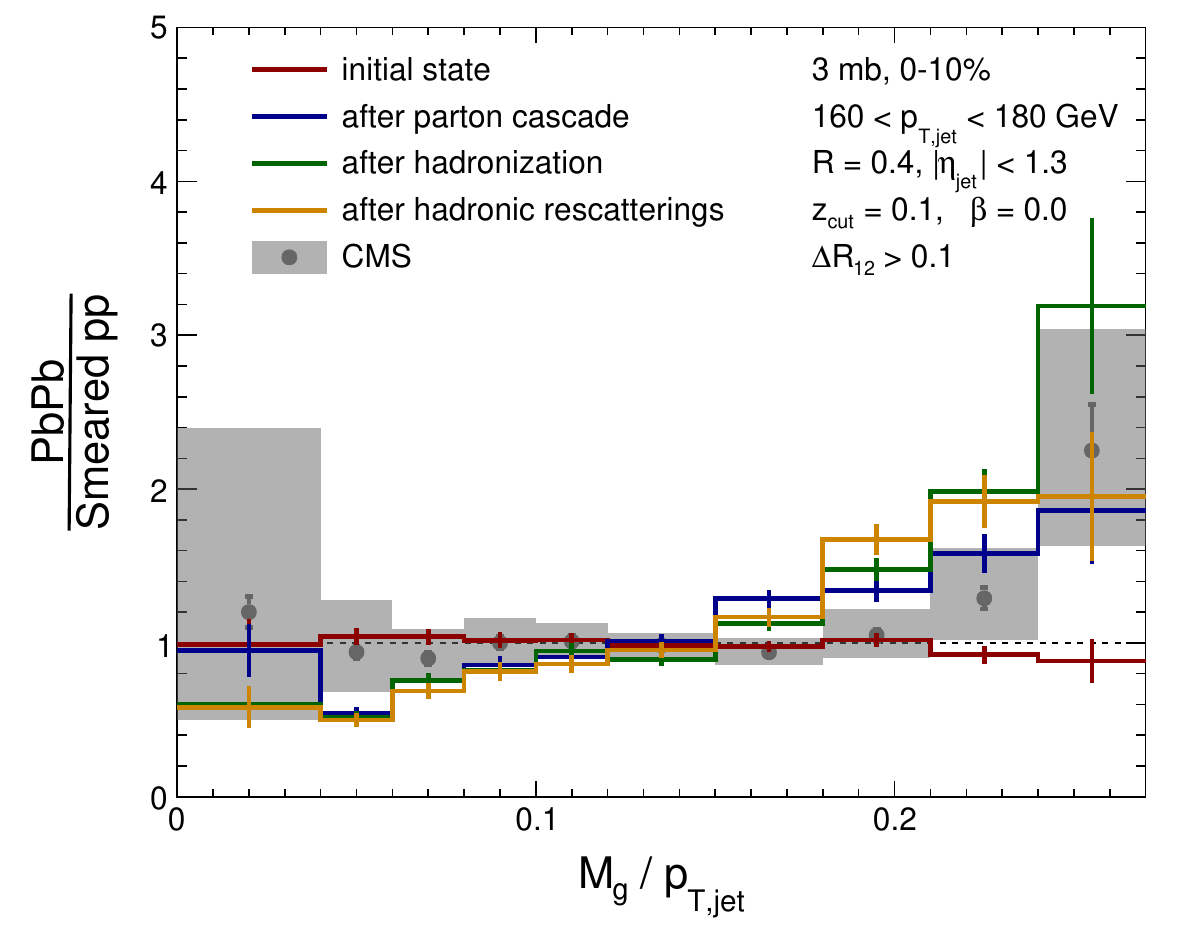}
    \includegraphics[width=0.42\textwidth]{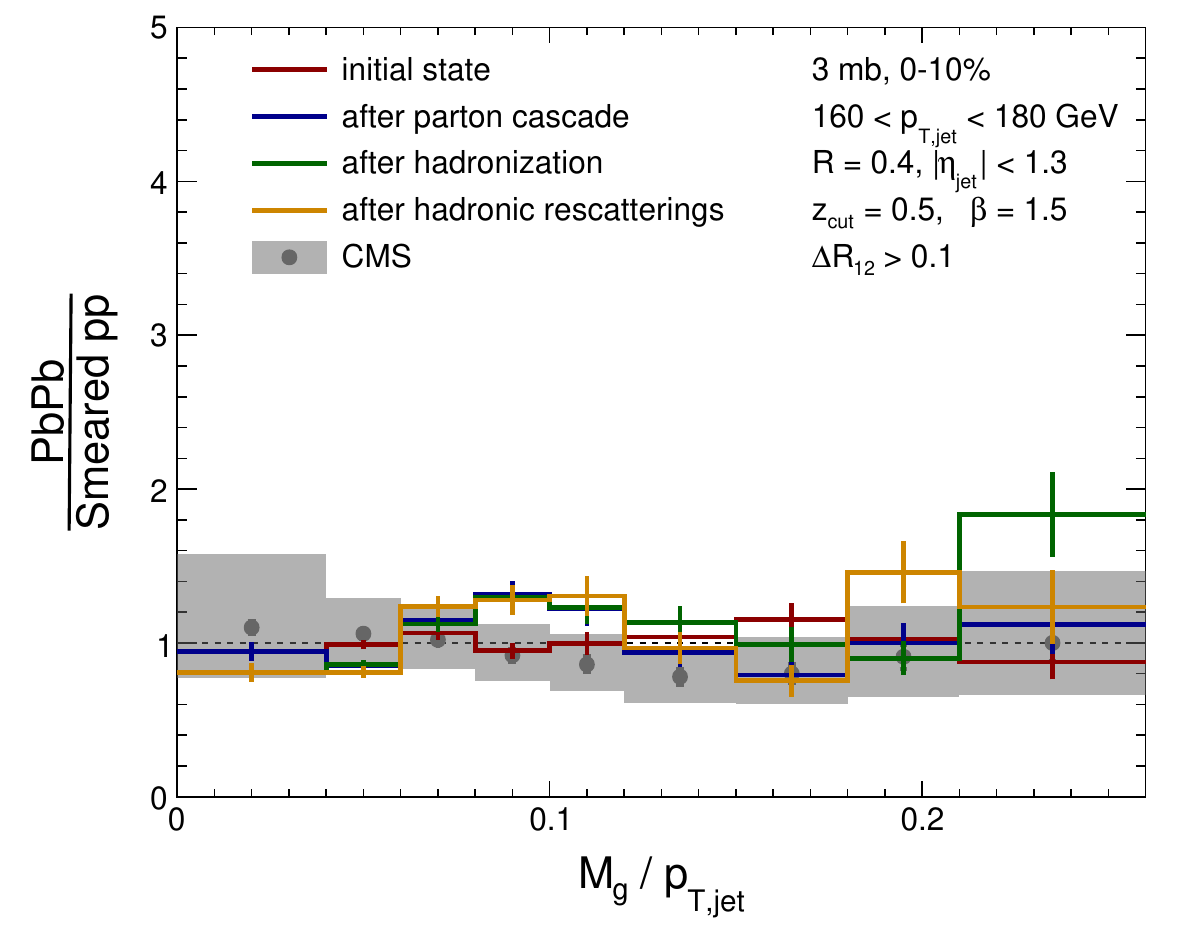}
    \includegraphics[width=0.42\textwidth]{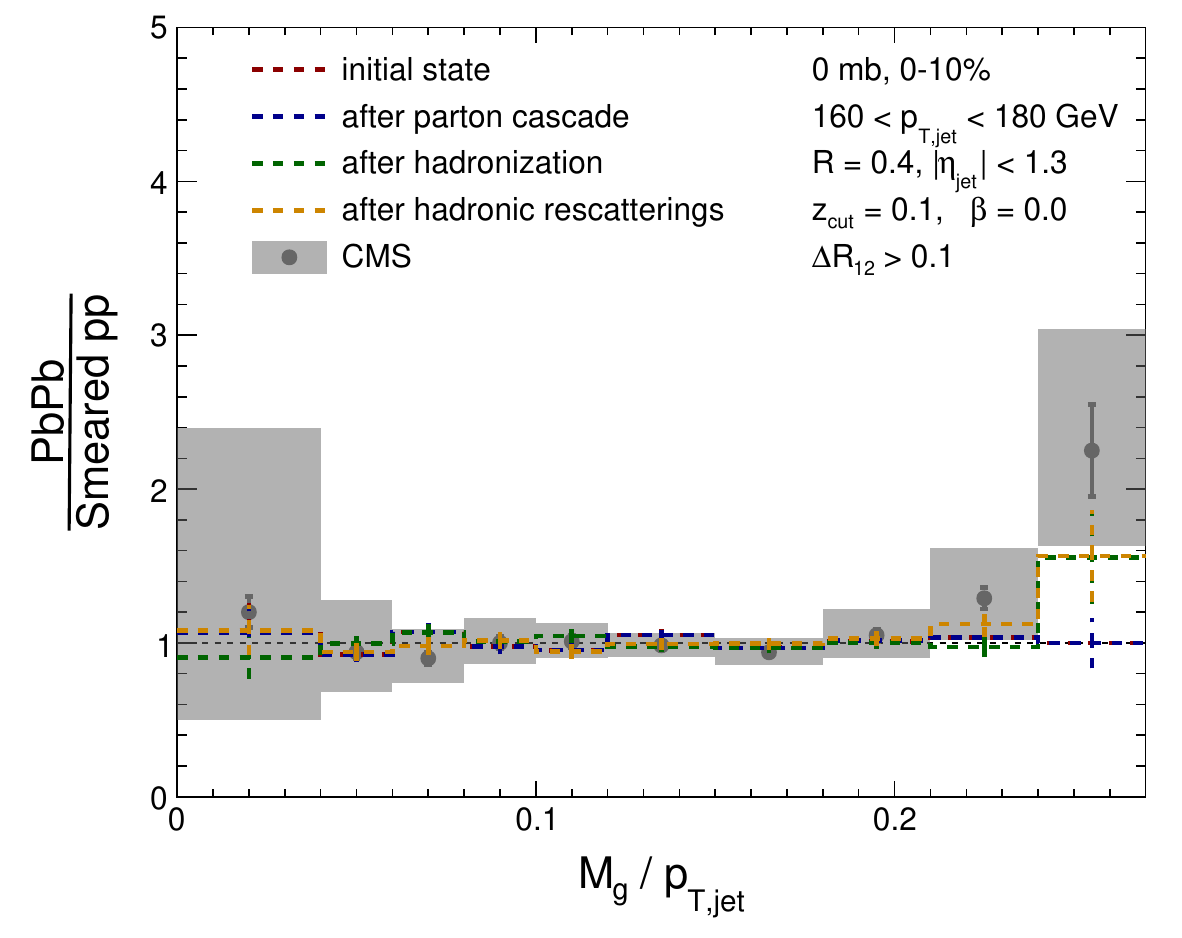}
    \includegraphics[width=0.42\textwidth]{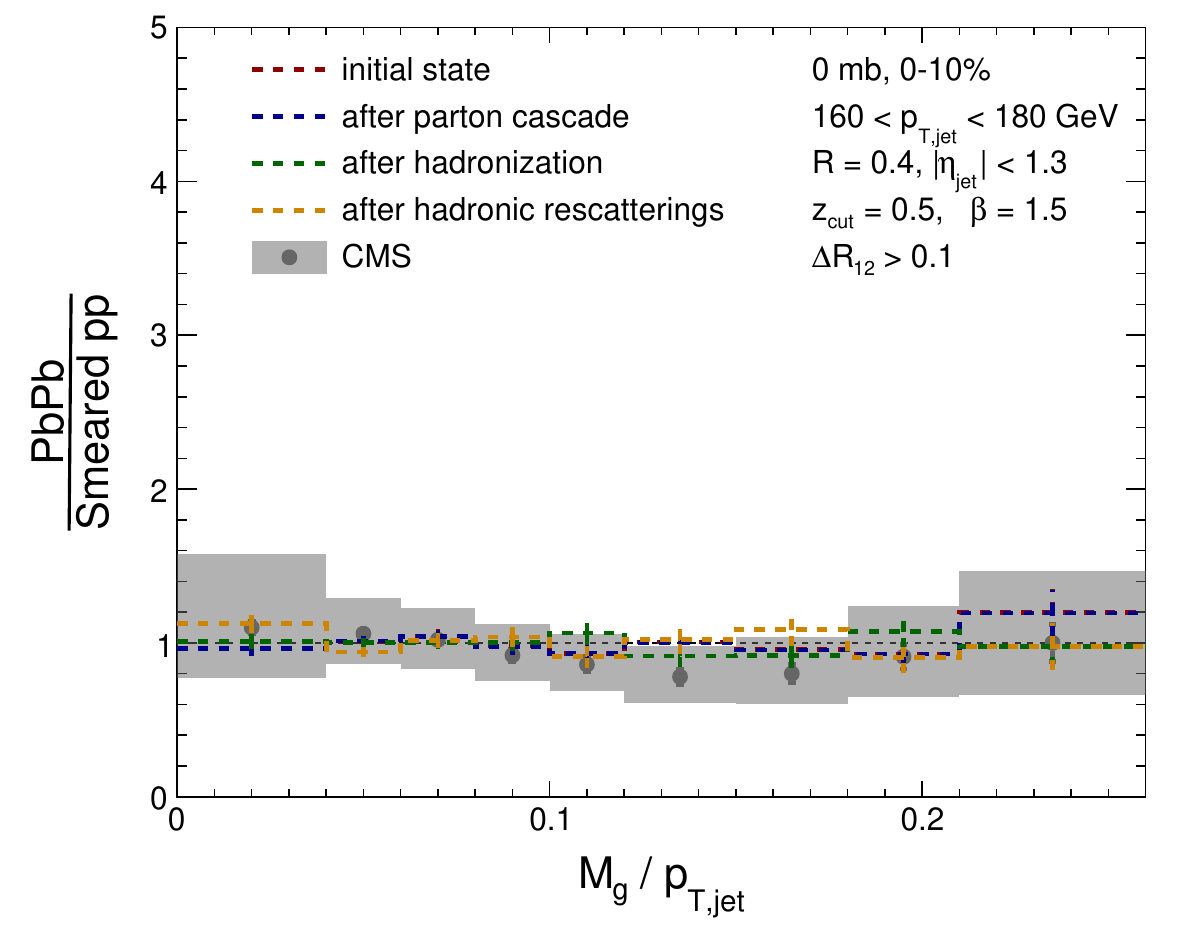}
    \caption{Ratios of the $M_g / p_{T,\text{jet}}$ distributions between 0-10\% PbPb and smeared $pp$ collisions in the jet transverse momentum range of $160 < p_{T,\text{jet}} < 180$~GeV within four dynamical evolution stages at $\sqrt{s_{NN}} = 5.02$~TeV. The Soft Drop grooming parameters are set to $z_\text{cut} = 0.1$, $\beta = 0.0$ (left panels) and $z_\text{cut} = 0.5$, $\beta = 1.5$ (right panels), with an angular cut of $\Delta R_{12} > 0.1$. Lines represent the AMPT model results with parton cross section of 3~mb (top panels) and 0~mb (bottom panels), while data points correspond to CMS measurements~\cite{CMS:2018fof}, with statistical uncertainties shown as error bars and systematic uncertainties as shaded bands.}
    \label{fig:PbPb-Mg-stage}
\end{figure*}

\section{Results and discussion}\label{sec:result}

\subsection{Jet splitting momentum fraction}\label{sec:result-zg}

We compare the $z_g$ distributions obtained using the first set of Soft Drop grooming parameters ($z_\text{cut} = 0.1$, $\beta = 0.0$) from the AMPT model with CMS measurements~\cite{CMS:2017qlm} at $\sqrt{s_{NN}} = 5.02$~TeV, as shown in Fig.~\ref{fig:pp-PbPb-Zg}. The left panel shows the comparison for $pp$ collisions in the jet transverse momentum range of $160 < p_{T,\text{jet}} < 180$~GeV. The AMPT results are in good agreement with the CMS data, providing a reliable baseline. The right panel presents the $z_g$ distributions for 0-10\% most central smeared $pp$ and PbPb collisions in the same $p_{T,\text{jet}}$ range. Here, the AMPT simulations are performed with partonic interactions of 3~mb. The centrality of smeared $pp$ events is determined based on the multiplicity distributions from the PbPb collisions without applying jet triggers. Compared to CMS data, the AMPT predictions show a slightly steeper slope.

Figure~\ref{fig:PbPb-Zg-cen-pT} displays the ratios of $z_g$ distributions between PbPb and smeared $pp$ collisions at $\sqrt{s_{NN}} = 5.02$~TeV, plotted as functions of centrality (left panel) and jet transverse momentum $p_{T,\text{jet}}$ (right panel). Two scenarios are considered: simulations with partonic interactions for 3~mb and without partonic interactions for 0~mb. In the left panel, for jets with $160 < p_{T,\text{jet}} < 180$~GeV, no modification is observed across all centrality intervals when partonic interactions are turned off. In contrast, simulations with partonic interactions exhibit negligible changes in peripheral PbPb collisions, while a slight enhancement for asymmetric splittings is observed in more central PbPb collisions. These results are qualitatively consistent with the CMS measurements within uncertainties. The right panel illustrates the $z_g$ modification as a function of jet transverse momentum $p_{T,\text{jet}}$ in 0-10\% most central collisions. For partonic interactions of 0~mb, the $z_g$ distributions remain largely unmodified across the $p_{T,\text{jet}}$ range of 140-200~GeV, with only minor deviations observed at higher jet momenta ($200 < p_{T,\text{jet}} < 250$~GeV). When partonic interactions are included, slight modifications appear throughout the full $p_{T,\text{jet}}$ range, indicating a weak sensitivity of $z_g$ to jet-medium interactions, particularly at low $p_{T,\text{jet}}$.

Furthermore, several theoretical studies have explored the behavior of the $z_g$ observable in the high jet transverse momentum regime ($p_{T,\text{jet}} > 140$~GeV), in comparison with CMS measurements~\cite{CMS:2017qlm}. Simulations using JEWEL with four-momentum subtraction method~\cite{Milhano:2017nzm}, which incorporates medium response effects, predict a mild shift of the $z_g$ distribution toward smaller values. Similarly, calculations based on the SCET incorporating Glauber gluon interactions~\cite{Chien:2016led} also reveal a slight enhancement in the probability of asymmetric splittings. At lower jet transverse momentum ($60 < p_{T,\text{jet}} < 100$~GeV), the ALICE experiment~\cite{ALICE:2021mqf} observes no significant modification of the $z_g$ distribution in PbPb collisions relative to $pp$ collisions, when using strong grooming settings ($z_\text{cut} = 0.2$, $\beta = 0.0$). These results are consistent with predictions from JETSCAPE~\cite{JETSCAPE:2023hqn}, which combines the MATTER model~\cite{Majumder:2013re} for high-virtuality parton evolution and the LBT model~\cite{He:2015pra,Luo:2023nsi} for low-virtuality in jet-medium interactions. Interestingly, the higher twist formalism~\cite{Chang:2017gkt}, which is based on the coherent energy loss assumption for the two split subjets, predicts a non-monotonic dependence of the $z_g$ modification on jet energy: the strongest modification occurs at intermediate jet energies, while both lower and higher energies exhibit weaker effects.

\subsection{Groomed jet mass}\label{sec:result-Mg}

We further investigate the groomed jet mass to the ungroomed jet transverse momentum, $M_g / p_{T,\text{jet}}$, using two sets of Soft Drop grooming parameters at $\sqrt{s_{NN}} = 5.02$~TeV, and compare our results with CMS data~\cite{CMS:2018fof}. Figure~\ref{fig:pp-PbPb-Mg} presents the $M_g / p_{T,\text{jet}}$ distributions in $pp$ collisions (top panels) and in both smeared $pp$ and PbPb collisions with partonic interactions of 3~mb (bottom panels), for jets in $p_{T,\text{jet}}$ range of 160-180~GeV. In the top panels, our simulations agree well with CMS results within uncertainties, providing a reliable baseline for further comparison. Additionally, the $M_g / p_{T,\text{jet}}$ distributions with the Soft Drop parameters $z_\text{cut} = 0.5$ and $\beta = 1.5$ (top right panel) appear steeper than those with $z_\text{cut} = 0.1$ and $\beta = 0.0$ (top left panel), reflecting a stronger grooming constraint, focused on the jet core. In the bottom left panel, corresponding to the first grooming condition, we observe a broadening of the $M_g / p_{T,\text{jet}}$ distribution in PbPb collisions compared to smeared $pp$ collisions. The smeared $pp$ result exhibits a minor deviation, whereas the PbPb result presents a slightly steeper slope in comparison with CMS data. For the second grooming condition, shown in the bottom right panel, the smeared $pp$ result remains consistent with CMS data within uncertainties, while the PbPb events display a deviation in the $M_g / p_{T,\text{jet}}$ distribution compared to CMS measurement.

Figure~\ref{fig:PbPb-Mg-cen} shows the centrality dependence of the ratio of $M_g / p_{T,\text{jet}}$ distributions between PbPb and smeared $pp$ collisions in the jet transverse momentum range of $160 < p_{T,\text{jet}} < 180$~GeV at $\sqrt{s_{NN}} = 5.02$~TeV, using two Soft Drop grooming parameters: a condition sensitive only to the energy fraction between subjets ($z_\text{cut} = 0.1$, $\beta = 0.0$) in the left panel, and a stronger grooming setup that emphasizes the jet core structure ($z_\text{cut} = 0.5$, $\beta = 1.5$) in the right panel. Simulations are performed both with (3~mb) and without (0~mb) partonic interactions, consistent with the approach taken in the $z_g$ analysis. In the left panel, when partonic interactions are turned off, the ratio remains flat across all centrality classes, with the exception of fluctuations in the highest $M_g / p_{T,\text{jet}}$ bin. However, when partonic interactions are included, a hint of enhancement emerges in the high $M_g / p_{T,\text{jet}}$ region, particularly in more central collisions, indicating medium-induced broadening of the groomed jet mass. In contrast, the stronger grooming condition shown in the right panel leads to no significant modification in any centrality intervals, regardless of the presence of partonic interactions. This confirms that the observed enhancement is predominantly associated with medium response located at larger angles from the jet axis.

To further investigate the sensitivity to jet transverse momentum, Fig.~\ref{fig:PbPb-Mg-pT} displays the $p_{T,\text{jet}}$ dependence of the $M_g / p_{T,\text{jet}}$ modification for 0-10\% most central PbPb collisions relative to smeared $pp$ collisions. For the first grooming condition (left panel), simulations without partonic interactions again show no modification, except for minor fluctuations in the highest $M_g / p_{T,\text{jet}}$ bin at intermediate $p_{T,\text{jet}}$. By contrast, when partonic interactions are included, a noticeable enhancement is observed in the low $p_{T,\text{jet}}$ region, consistent with CMS measurements. This suggests that elastic interactions between jet and medium partons contribute to an increase in the groomed jet mass. No significant changes are observed across all $p_{T,\text{jet}}$ intervals under the stronger grooming condition (right panel), further confirming that medium-induced modifications are primarily attributed to large-angle scattering.

To elucidate the origin of these modifications, we further analyze the evolution of $M_g / p_{T,\text{jet}}$ at four dynamic stages in AMPT model: initial state, after parton cascade, after hadronization, and after hadronic rescatterings, as shown in Fig.~\ref{fig:PbPb-Mg-stage}. The analysis focuses on jets with $160 < p_{T,\text{jet}} < 180$~GeV in 0-10\% most central PbPb and smeared $pp$ events. For the first grooming condition (top left panel), no modification is observed at initial state, while a pronounced enhancement in the $M_g / p_{T,\text{jet}}$ distribution emerges immediately after parton cascade stage (3~mb), confirming that jet-medium interactions are the primary drivers of the modification. Hadronization and hadronic rescatterings stages retain the enhancement, but their additional effects are minor, consistent with the ability of grooming procedure to suppress non-perturbative contributions. In contrast, for simulations without partonic interactions (bottom left panel), no noticeable change is seen across any stage, reaffirming the absence of jet quenching effects. Results using the second grooming condition are shown in the right panels, where only minor deviations are observed, further emphasizing that the groomed jet mass is sensitive to medium response located at larger angles from the jet axis.

In a word, our analysis demonstrates that jet-medium interactions lead to a significant enhancement in the groomed jet mass. This effect has been investigated in several theoretical frameworks. JEWEL simulations~\cite{Milhano:2022kzx} fail to describe the smeared $pp$ and PbPb results, in comparison with CMS data~\cite{CMS:2018fof}. Nevertheless, medium response in JEWEL still leads to a noticeable enhancement at large $M_g / p_{T,\text{jet}}$ values~\cite{CMS:2018fof}. Similarly, LBT model~\cite{Luo:2021iay} predicts a substantial enhancement in the tail of the groomed jet mass distribution, due to medium response. When comparing with ALICE data~\cite{ALICE:2024jtb}, Hybrid model~\cite{Casalderrey-Solana:2014bpa}, based on a strongly coupled AdS/CFT framework, shows that the elastic Moli\`{e}re scattering~\cite{DEramo:2018eoy} generates large-angle jet constituents and broadens the groomed jet mass distributions. These results collectively highlight the strong sensitivity of the groomed jet mass to jet-medium interactions, making it a powerful observable for probing QGP properties.

\section{Summary}\label{sec:summary}

In this study, we explored jet substructure observables in $pp$ and PbPb collisions at $\sqrt{s_{NN}} = 5.02$~TeV using the AMPT model with the string melting mechanism. Background effects in relativistic heavy-ion collisions were mitigated using the constituent subtraction method for both PbPb and smeared $pp$ events. The analysis focused on two groomed jet observables reconstructed via the Soft Drop algorithm: the jet splitting momentum fraction ($z_g$) and the groomed jet mass to the ungroomed jet transverse momentum ($M_g / p_{T,\text{jet}}$). Two sets of grooming parameters were employed to probe the sensitivity of these observables to jet-medium interactions. The simulation results were compared with CMS data.

Our simulations reproduce the $z_g$ and $M_g / p_{T,\text{jet}}$ distributions in $pp$ collisions, establishing the AMPT model as a reliable baseline. With the Soft Drop parameters $z_\text{cut} = 0.1$ and $\beta = 0.0$, a slight enhancement of asymmetric splittings in the $z_g$ distribution is observed in central PbPb collisions relative to smeared $pp$ events, in qualitative agreement with CMS measurements. In contrast, a more significant modification is found in the $M_g / p_{T,\text{jet}}$ distribution, particularly in more central events and at lower jet transverse momentum. By tracing the dynamical evolution of jet substructure through the different stages in the AMPT model, we find that this enhancement originates primarily from the parton cascade stage, where elastic scatterings between the jet and medium partons dominate. Contributions from hadronization and hadronic rescatterings are strongly suppressed by the grooming procedure. Furthermore, simulations with the stronger grooming constraint ($z_\text{cut} = 0.5$, $\beta = 1.5$) show no significant changes, underscoring that the medium-induced modifications are predominantly associated with large-angle scattering. Moreover, previous AMPT studies~\cite{Gao:2016ldo,Luo:2021hoo} have emphasized the role of elastic interactions in transporting energy from the hard jet core to larger angles. Incorporating inelastic radiative energy loss processes into the parton cascade is expected to enhance soft gluon radiation and result in more small-angle splittings, as reflected in jet splitting radius ($r_g$)~\cite{ALICE:2021mqf,JETSCAPE:2023hqn} during jet-medium interactions. This represents a crucial direction for the future development of the AMPT framework.

\begin{acknowledgments}
We thank Prof. Bin Wu for helpful discussions, and Dr. Chen Zhong for maintaining the high-quality performance of the Fudan supercomputing platform for nuclear physics. This work is supported by the National Natural Science Foundation of China under Grants No.12147101, No. 12325507, the National Key Research and Development Program of China under Grant No. 2022YFA1604900, and the Guangdong Major Project of Basic and Applied Basic Research under Grant No. 2020B0301030008. X.D.~is supported by the China Scholarship Council under Grant No. 202306100165; by European Research Council project ERC-2018-ADG-835105 YoctoLHC; by Maria de Maeztu grant CEX2023-001318-M and by project PID2023-152762NB-I00 both funded by MCIN/AEI/10.13039/-501100011033; from the Xunta de Galicia (CIGUS Network of Research Centres) and the European Union.
\end{acknowledgments}

\bibliographystyle{apsrev4-2}
\bibliography{ref.bib}

\end{document}